\documentclass[12pt]{article}
\title{Instanton--anti--instanton molecule with non--zero--modes of quarks
included}
\author{ A.G.Zubkov,
O.V.Dubasov, B.O.Kerbikov,
 \\ Institute of
Theoretical and Experimental Physics\\ 117218, Moscow, B.Cheremushkinskaya
25, Russia\\
E-mail:borisk@heron.itep.ru}
\date{}
 \newcommand{\be}{\begin{equation}}
\newcommand{\ee}{\end{equation}}

\begin{document}
\maketitle

\begin{abstract}

 A straightforward algebraic derivation is carried out of light quark
propagator and effective action in the instanton--anti--instanton molecule.
Exact expressions are obtained which contain  contributions of all quark
modes. Possible implications of the results include chiral phase transition
and correlation functions of mesons and baryons.

\end{abstract}

\section{Introduction}

The theory of light quarks in the instanton vacuum has a long history -- see
e.g. review papers [1-3]. Recently a new impetus to the instanton liquid
model\footnote{Sometimes also called instanton gas model} has been provided
by lattice measurements of quark and gluon correlators in coordinate space
[4] showing the important role of instantons in forming hadronic wave
functions and mass spectra. The standard theory of the instanton vacuum is
based on the assumption of quark zero--modes dominance. This assumption was
questioned in [5], and it was argued that non-zero--modes of each instanton
play an important role in addition to zero--modes. To our knowledge it was
the   first systematic treatment of higher modes although their role in
specific  effects and within various approximations were considered in some
earlier papers [6-8].

Here we wish to present a complete treatment with all modes  included of the
so--called instanton--anti--instanton molecule, i.e. the $I\bar I$
pair "bound" through light quarks. This object plays an important
role in the instanton model of the QCD vacuum. According to the
standard scenario the chiral symmetry restoration occurs due to the
onset of the molecular component in the instanton--anti--instanton
ensemble [9-15]. The analysis of  chiral phase transition is  not the
purpose of the present paper. Our main objective is the detailed
calculation of the molecule including quark propagator and effective
action. Tooled with these results one can reexamine the chiral
symmetry restoration in more details. Another  even more immediate
implication is the  study of interactions in mesons and baryons induced by
correlated $I\bar I$ pairs [14]. The corresponding effective
Lagrangian is derived from the quark propagator. As we shall see the
quark propagator in the   quantity in which the effects of
non--zero--modes are mostly propounced.

The systematic study of $I\bar I$  molecule was performed in 1979 by C.Lee
and W.A.Bardeen [16]. The propagator and  the effective action were
constructed making use of the power series in two independent parameters
$m\rho$ and $\rho/d$, where $m$ is the quark dynamical mass, $\rho$ is the
size of (anti--) instanton, $d$-- the distance between the two centers. In
the propagator the interference terms between zero--modes and continuum were
taken into account in the leading order with respect to the above parameters.
Using this propagator the quark effective action was constructed.
Surprisingly enough it contains only zero--modes overlap. In other words it
was shown that the propagator is more affected by non-zero--modes than the
effective action. The overlap of zero--modes in the effective action
describes logarithmic attraction between instanton and anti--instanton.

The revival of the interest to the role of non-zero--modes on one
hand and to the $I\bar I$ molecule on the other calls for the
reexamination of the problem. Below we develop an approach different
from that of [16]. In its spirit our formalism is close to the
multiple scattering theory [17,13], or scattering theory in a system
with confinement [18]. Along these lines we have found the exact
solution of the problem with all modes included. The final answers
for the quark propagator and effective action are given in terms of
matrix elements involving individual (anti--) instanton zero--modes
and individual propagators with zero--modes deleted. Unfortunately
the corresponding integrals may not be evaluated analytically and we
estimate them by the orders of magnitudes with respect to the
parameters $\rho/d$ and $m\rho$. In this way we present the
non--zero--modes corrections to the effective action evaluated by
Lee and Bardeen [16]. To get accurate results one has to perform
numerical calculations which is beyond the scope of this paper.

 The outline of this work is as follows. In Section 2 we introduce notations
and derive the "two--centers" equation for the quark propagator in
$I\bar I$
 molecule. Section 3 is devoted to the evaluation of the quark propagator in
$I\bar I$
molecule. We present an exact expression for the propagator and an
approximate form in which  higher modes contribution is accounted for
in the leading order. We argue that the  interference of zero modes
with continuum is  generally of the same order as zero  modes
contribution. Section 4 deals with fermionic effective action. Again
we find an exact answer with the contributions of all modes included.
In Section 5 we discuss estimates for the propagator and for the
effective action based on the existence of two small parameters --
the ratio of the insatanton size to the size of the molecule and the
quark current mass. Finally in Section  6 we discuss our results and
suggest various ways of their implementations.

\section{Basic definitions and the general approach }

As usual in the instanton physics we work in Euclidean metric. Also
in line with the common lore we use the superposition ansatz for the
$I\bar I$
system
\be
A_{\mu}(x) =\sum_{i=1,2} A_\mu^{(i)}(x-R_i)= A^I_\mu(x)+A_\mu^{\bar I}(x-d),
\ee
where $i=1$ corresponds to $I$ placed at the origin $(R_1\equiv R_I=0),~~i=2$
corresponds to $\bar I$ placed at $R_2\equiv R_{\bar I}=d$. In singular gauge
we can write
\be
gA_\mu^I(x)=\frac{\bar\eta_{a\mu\nu}x_\nu\rho^2\Omega^+_I\tau_a\Omega_I}
{x^2(x^2+\rho^2)},
\ee
where $\rho$ and $\Omega_I$ are the size and color orientation of the
instanton. Anti-instanton is assumed to have the same size $\rho$ (this is
not crucial), and its field follows from (2) though the substitution $\bar
\eta\to\eta$ and $x\to (x-d)$. Next we introduce the free propagator $S_0$,
the single (anti--) instanton propagar $S_i$ with $i=1$ for $I$ and $i=2$ for
$\bar I$, and the total propagator $S$ in the field (1):
$$
S^{-1}_0=(-i\hat\partial - i m),
$$
$$
S^{-1}_i=(-i\hat\partial -  g\hat A^{(i)}-i m),
$$
\be
S^{-1}=(-i\hat\partial -  g\hat A-i m).
\ee
 Note that our free propagator $S_0$ includes the current quark mass $m$
which is not the case in Lee and Bardeen paper [16].

The instanton propagator has a formal  spectral representation over normal
modes
\be
S_I=\sum^\infty_{n=0}\frac{|u^I_n><u^I_n|}{\lambda_n-im},
\ee
with eigenfunctions $|u^I_n>$ satisfying the equation
\be
(-i\hat \partial-g\hat A^I)|u^I_n>=\lambda_n|u^I_n>.
\ee
Of special role are zero--modes with $\lambda_0=0$ which have negative
chirality
\be
\gamma_5|u^I>=-|u^I>,
\ee
where index $n=0$ is omitted. Anti--instanton zero--modes $|u^{\bar I}>$ have
positive chirality. Zero--mode has the following explicit form
$$
(u^I(x))_{ci}=\frac{\rho}{\pi(2x^2)^{1/2}(x^2+\rho^2)^{3/2}}
[(\frac{1-\gamma_5}{2})\hat x]_{ij}
 \varepsilon_{jc'}\Omega_{c' c}^+=
$$
\be
=\frac{\rho}{\pi(2x^2)^{1/2}(x^2+\rho^2)^{3/2}}
(\tau_x)_{ij}
 \varepsilon_{jc'}\Omega_{c' c}^+,
\ee
where $c$ and $i$ are color and Lorentz indices, $\varepsilon$ is
antisymmetric tensor,
$\tau_x=\tau_\mu x_\mu,~\tau_\mu=(\vec \tau,i),~~
~\tau_\mu^+=(\vec \tau,-i)$ with $\vec \tau$ being the Pauli matrix. For
anti--instanton $x$ is replaced by $(x-d),~~ (1-\gamma_5)\to (1+\gamma_5),
\tau_x\to \tau^+_{x-d}$.

The non-zero--modes part of the instanton propagator (3-4) is known
explicitly in the chiral limit $m\to 0$ [19]  and given by [19,6,7]
$$
S'_I(x,y)=\sum_{n>0}\frac{u^I_n(x)(u^I_n(y))^+}{\lambda_n}=
$$
\be
=G_0(x-y)(1+\frac{\rho^2}{x^2y^2} \tau_x\tau^+_y) (\pi_x\pi_y)^{-1/2}-
\ee
$$
-\frac{\rho^2(\pi_x\pi_y)^{-1/2}}{4\pi^2x^2y^2(x-y)^2}
\tau_x[\frac{\hat\tau^+\tau_{x-y}}{x^2+\rho^2}
(\frac{1+\gamma_5}{2}) +
\frac{\tau^+_{x-y}\hat\tau}{y^2+\rho^2}
(\frac{1-\gamma_5}{2})]\tau_y,
$$
where $G_0(x-y)$ is the free propagator of massless fermion (not to
be confused
with $S_0$ given by (3) in which the quark current mass is included)
\be
G_0(z)=-\frac{i}{2\pi^2}\frac{\hat z}{z^4},
\ee
and
\be
\pi_z=1+\rho^2/z^2.
\ee
The propagator $S'_I(x,y)$ obeys the Green's function equation with $m=0$
and zero--mode deleted
\be
(-  i\hat \partial-g\hat A^I)S'_I={\bf 1\!I}
-|u^I><u^I|.
\ee
Mass corrections of the order $0(m)$  to $S'_I$ can be easily found
from the spectral representation (4) rewritten in the form
$$
S_I=\frac{|u^I>< u^I|}{-im}+\sum_{n>0}\frac{\lambda_n}{\lambda^2_n+m^2}
|u^I_n><u^I_n|+im\sum_{n>0}\frac{|u^I_n><u^I_n|}{\lambda^2_n+m^2}.
$$
In the chiral limit the second term in the right--hand side yields $S'_I$
(see the first line in (8)). To get  $0(m)$ correction from the third term we
note that due to the orthogonality of eigenmodes
$$
\sum_{n>0}\frac{|u^I_n><
u^I_n|}{\lambda^2_n}=(\sum_{n>0}
\frac{|u^I_n><u^I_n|}{\lambda_n})^2=
(S'_I)^2.
$$

Thus the quark propagator in the single instanton field up to order $m^2$
terms has form [6]
$$
S_I(x,y)=\frac{u^I(x)(u^I(y))^+}{-im}+
S'_I(x,y)+
$$
\be
+im\int dz S'_I(x,z)S'_I(z,y).
\ee

We note in passing  that the instanton liquid model (see [2,3] and
references therein) relies on the following approximation for the quark
propagator in the single instanton field
\be
S_I=\frac{|u^I><u^I|}{-im} +(-i\hat \partial)^{-1}.
\ee
The validity of this ansatz in instanton liquid model was discussed in [5].
The situation for the $I\bar I$ molecule will be elucidated in what follows.

For  further purposes it will be convenient to recast (12) into the following
form
\be
S_I=\frac{|u^I><u^I|}{-im} +S_0+\Delta_I,
\ee
so that
$$
\Delta_I\equiv \Delta_1=S'_I-S_0+im (S'_I)^2=
(S'_I-G_0)+im[(S'_I)^2-G^2_0]\equiv
$$
\be
\equiv \delta_1+im[(S'_I)^2-G^2_0],
\ee
where $G_0$ is given by (9). For $\bar I$ similar notations are introduced
with $\Delta_{\bar I}\equiv \Delta_2$.

Now we derive the two--centers equation for the propagator $S$ defined by
(3). It follows straightforwardly from the superposition ansatz (1) and the
definitions (3). Namely one can write
$$
S=\frac{1}{S_1^{-1}+S_2^{-1}-S_0^{-1}}=
S_2\frac{1}{S_1(S^{-1}_1+S_2^{-1}-S_0^{-1})S_2}S_1=
$$
\be
=S_2\frac{1}{S_0-(S_1-S_0)S_0^{-1}(S_2-S_0)}S_1=S_1
\frac{1}{S_0-(S_2-S_0)S_0^{-1}(S_1-S_0)}S_2.
\ee
Generalization for more than two centers may be found in [13]. Next task is
the actual calculation of (16), i.e. the invertion of the operator in the
denominator. This will be done exactly without resorting to the
approximations of the type
$$
S\simeq S_0+\sum_i(S_i-S_0).
$$

\section{Quark propagation function in the molecule}

Using (14) the propagator (16) may be rewritten as
\be
S=S_2Z^{-1}S_1,
\ee
$$
Z=S_0-\Delta_1S_0^{-1}\Delta_2-
\frac{|u^1><u^1|}{-im}
S^{-1}_0\Delta_2- \Delta_1S_0^{-1}
\frac{|u^2><u^2|}{-im}+
$$
\be
+|u^1>\frac{V_{12}}{m^2}<u^2|,
\ee
\be
V_{12}  =<u^1|S_0^{-1}|u^2>=<u^1|-i\hat\partial |u^2>=
<u^1|g\hat A^{(2)}|u^2>.
\ee

Performing calculations one may use any of the two equivalent expressions
(16) for  $S$. At intermediate stages of calcutions the $1\leftrightarrow 2$
permutation symmetry is lost as in (18). In final answers this symmetry is
restored.

To find $Z^{-1}$ we will use the technique based on the relation
\be
\frac{1}{A+|\zeta>a<\eta|}=\frac{1}{A}-\frac{1}{A}|\zeta>\frac{a}{1+<\eta|
\frac{a}{A}|\zeta>}<\eta|\frac{1}{A},
\ee
with $A$ being an operator and $a$ -- $c$-number. Applying it to (18) one
gets
$$
Z^{-1}=\frac{1}{W+|u^1>\frac{V_{12}}{m^2}<u^2|}=
$$
\be
=\frac{1}{W}-\frac{1}{W}|u^1>\frac{V_{12}}{m^2+V_{12}<u^2|W^{-1}|u^1>}
<u^2|\frac{1}{W},
\ee
\be
W=K-\frac{|u^1><u^1|}{-im} S_0^{-1}\Delta_2-\Delta_1 S_0^{-1}
\frac{|u^2><u^2|}{-im},
\ee
\be
K=S_0-\Delta_1S_0^{-1}\Delta_2.
\ee

To find $W^{-1}$ one has to apply equation (20) repeatedly twice.
After some straightforward but tedious work of collecting all terms
in (21) we arrive at
 $$ Z^{-1}=K^{-1}+\frac{im+\alpha}{D}
K^{-1}|u^1><u^1| R_2K^{-1}+
\frac{im+\beta}{D}
K^{-1}R_1|u^2><u^2|K^{-1}-
$$
\be
\frac{V_{12}+A_{12}}{D}
K^{-1}|u^1><u^2|K^{-1}-
\frac{V_{21}+A_{21}}{D}
K^{-1}R_1|u^2><u^1|R_2K^{-1},
\ee
where
$$
\alpha=<u^2|K^{-1}\Delta_1 S_0^{-1}|u^2>,
$$
$$
\beta=<u^1|S^{-1}_0\Delta_2 K^{-1}|u^1>,
$$
$$
V_{12}=<u^1|S^{-1}_0|u^2>,
V_{21}=<u^2|S^{-1}_0|u^1>,
$$
\be
A_{12}=<u^1|S^{-1}_0\Delta_2 K^{-1}\Delta_1S_0^{-1}|u^2>,
\ee
   $$
A_{21}=<u^2|S^{-1}_0\Delta_1 \tilde K^{-1}\Delta_2S_0^{-1}|u^1>,
$$
  $$
K=S_0-\Delta_1S_0^{-1}\Delta_2,~~
\tilde K=
S_0-\Delta_2S_0^{-1}\Delta_1,
$$
$$
D=m^2+(V_{12}+A_{12})(V_{21}+A_{21})-im (\alpha+\beta)-\alpha\beta.
$$
$$
R_i|u^j>=\Delta_iS^{-1}_0|u^j>, ~~<u^j|R_i=<u^j|S^{-1}_0\Delta_i,~~ i,j=1,2.
$$
The final step is to insert (24) into (17) and use (14) for $S_i$. This
yields the following expression for the propagator
$$
S=\frac{1}{2}(S_0+\Delta_2)K^{-1}(S_0+\Delta_1)+\frac{1}{2}(S_0+\Delta_1)
\tilde K^{-1}(S_0+\Delta_2)+
$$
$$
+\frac{im+\alpha}{D}(S_0+\Delta_2)K^{-1}|u^1><u^1|\tilde
K^{-1}(S_0+\Delta_2)+
$$
$$
+\frac{im+\beta}{D}(S_0+\Delta_1)\tilde K^{-1}|u^2><u^2|
K^{-1}(S_0+\Delta_1)-
$$
$$
-\frac{V_{12}+A_{12}}{D}(S_0+\Delta_2) K^{-1}|u^1><u^2|
K^{-1}(S_0+\Delta_1)-
$$
\be
-\frac{V_{21}+A_{21}}{D}(S_0+\Delta_1)\tilde K^{-1}|u^2><u^1|
\tilde K^{-1}(S_0+\Delta_2).
\ee
This expression is an exact one. It relies only on the standard
superposition assumption (1) for the molecular field and on the formal
representation (14) for the individual instanton propagator. All individual
(anti--) instanton modes are included. No any small parameter like $\rho/d$
or $m\rho$ was used in the derivation. To get from (26) an insight into the
physical picture one should either perform numerical calculations or resort
to approximations and estimates. This last strategy will be used in what
follows.

Keeping in (26) only linear terms in $\Delta_i$ one obtains
$$
S\simeq S_0+\Delta_1+\Delta_2+\frac{im+\alpha}{d}(1+R_2)|u^1><u^1|(1+R_2)+
$$
$$
+\frac{im+\beta}{d}(1+R_1)|u^2><u^2|(1+R_1)-
$$
\be
-\frac{V_{12}}{d}(1+R_2)|u^1><u^2|(1+R_1)-
\frac{V_{21}}{d}(1+R_1)|u^2><u^1|(1+R_2),
\ee
where
\be
d=m^2+V_{12}V_{21}-im(\alpha+\beta).
\ee
This approximation coincides with the result of Lee and Bardeen [16]. In the
chiral
limit, more precisely when $m^2\ll V_{12}V_{21}$, one gets even
simpler expression $$ S\simeq
S_0+\Delta_1+\Delta_2-V^{-1}_{21}(1+R_2)|u^1><u^2|(1+R_1)- $$ \be
-V^{-1}_{12}(1+R_1)|u^2><u^1|(1+R_2).
\ee
It is tempting to simplify (29) further and neglect $\Delta_i$ and
$R_i$ (see e.g.[3,14]). However such approximation is not justified
since as it was already mentioned in [16] and will be discussed below
in Section 5 all terms in (29) are in general of the same order of
magnitude.

\section{Fermionic effective action}

We define the quark effective action in $I\bar I$ molecule as
\be
exp\{\Gamma_{eff}\} =\frac{det S^{-1}}{det S^{-1}_0}
\ee

We leave aside at least temporarily the questions of normalization and
regularization -- see, e.g. [20-22]. From (16) we get
\be
det S^{-1}=det S^{-1}_1\cdot det S^{-1}_2\cdot detZ
\ee
where the quantitity $Z$ is given by Eq. (18) of the previous
section. To calculate $det Z$ use can be made of an equation similar
to (20) \be det (A+|\zeta>a<\eta|)=det A\cdot
(1+<\eta|\frac{a}{A}|\zeta>).  \ee It is a simple but tedious task
to apply (31) repeatedly to (18). In fact no new calculations has be
done since the needed quantity essentially coincides with $D$
defined by (24-25). Precisely
 \be
  det Z= det S_0\cdot det
(1-S^{-1}_0\Delta_1 S^{-1}_0 \Delta_2) \frac{D}{m^2}.
\ee
Combining (30), (31) and (33) one obtains
\be
exp\{\Gamma_{eff}\} =(\frac{det S^{-1}_1}{m det S^{-1}_0})
(\frac{det S^{-1}_2}{m det S^{-1}_0}) det
(1-S^{-1}_0\Delta_1S^{-1}_0\Delta_2)D
\ee
If we now perform standard regularization and normalization procedures
[20-22] the final result for the effective action would be
$$
\Gamma_{eff}=
\Gamma_{eff}^I+
\Gamma_{eff}^{\bar I}+ ln det_{reg}(1-
S^{-1}_0\Delta_1 S^{-1}_0 \Delta_2^{-1})+
$$
 \be
+N_f ln[m^2+(V_{12}+A_{12})(V_{21}+A_{21})-im
(\alpha+\beta)-\alpha\beta].
 \ee
  This result is an exact one and
does not rely upon any approximation except for the superposition
ansatz (1). The first two terms describe effective actions of
individual (anti--) instantons. The third term corresponds to
no--zero modes contribution. Approximated by $\simeq -tr (S_0^{-1}
\Delta_1 S^{-1}_0 \Delta_2)$ it exhibits exponential damping as compared  to
the last one. This last term is of the main interest. It consists of the
individual zero modes contribution and the interference between zero modes
and continuum. The well known approximation to  this term is
$N_fln(m^2+V_{12}V_{21})$ [16]. In the next section we shall estimate
corrections to this expression using the exact result (35).

\section{Approximations and estimates}

In Section 3 the fermion propagator (26) has been determined and in the
previous section the fermion induced effective action (35) has been found.
However both expressions are not easy to handle due to high--dimensional
integrals which cannot be evaluated analytically. Therefore in the present
section we shall resort to estimates based on the existence of two
independent parameters supposed to be small. These are $\rho/d$ and $m\rho$,
where $\rho$ is the (anti--) instanton size, $d$-- the $I\bar I$ separation,
$m$ - the current mass of the quark. The widely accepted values of $\rho$ and
$d$ are: $\rho\simeq 0.3 fm, d\simeq 1 fm$ [1-4]. The key quantities entering
into most integrals (matrix elements) are $\Delta_i$ given by (14-15).
According to (15) $\Delta_i$ is splitted into mass independent part
$\delta_i(x,y)=S'_i(x,y)-G_0(x,y)$ and a contributions
proportional to $m$. Therefore corrections with
respect to $\rho/d$ and $m\rho$ are independent.

Let us start by providing arguments in favor of the statement made after
equation (29) for the propagator, namely that zero modes interference with
continuum is of the same order as zero modes product.

Consider the situation when both $x$ and $y$ are close to the instanton,
$x\sim y\sim \rho$. Then using (7) for zero modes one arrives at the
following estimate for zero modes product
\be
<x|u^2><u^1|y>\sim \rho^{-1}d^{-3}.
\ee
In the same region the interference term $R_1|u^2><u^1|$ may be estimated
using (8) and (9). This yields
$$
<x|R_1|u^2><u^1|y>\simeq(\int dz \delta_1 (x,z) \partial_z
u^2(z-d))<u^1|y>\sim
$$
\be
\sim(\rho^4 d^{-3}\rho^{-3})\rho^{-2}\sim \rho^{-1} d^{-3}.
\ee
In (37) we neglected mass corrections and substituted $\Delta_1$ by
$\delta_1$ and $S_0^{-1}$ by $\partial_z$, then performed
integration over the shell of size $\rho$ around $z\sim d$ which
resulted in $\rho^4$ volume. Also unjustified (at least in certain
regions) is the neglect of $\Delta_i$ as compared to $S_0$ in (29).
For  example when $x\sim y\sim \rho$  and  $(x-y)\sim\rho$ both $S_0$
and $\Delta_1$ are of  the order $\rho^{-3}$. On the other hand when
$x\sim\rho$ and $y\sim d$ zero modes product dominates over
interference.

From the above example it is clear that to make estimates one should
consider different regions of integration, typically the "near"
region $x\sim y\sim \rho$, the "far" region  $x\sim y\sim d$, and the
"mixed" region $x\sim \rho,y\sim d$. Looking for the estimate with
repect to $\rho/d$ one substitutes $S^{-1}_0$ either by $1/d$, or by
$1/\rho$. Attention should be paid to chirality arguments, namely the
matrix elements $<u^1|\gamma_{\mu}\gamma_{\nu}...\gamma_{\rho}|u^2>$
and $<u^1|\gamma_{\mu}\gamma_{\nu}...\gamma_{\rho}|u^1>$ are
respectively nonzero for odd and even number of $\gamma$ matrices.

Consider now the essential part of the effective action (35)
\be
\tilde \Gamma_{eff} = N_f ln [m^2+(V_{12}+A_{12})(V_{21}+A_{21})-im
(\alpha+\beta)-\alpha\beta]
\ee
The overlaps of zero modes $V_{ij}(i,j=1,2)$ are of the order
$V_{ij}\sim \rho^2 d^{-3}$ [1-3,16]. The leading correction to (38)
in $\rho/d$ stems from $A_{ij}$ while that in $m\rho$ arises from
$\alpha $ and $\beta$. Let us begin with the correction in
$\rho/d$. For definiteness consider $A_{12}$. We are interested in the
leading  order correction and therefore can substitute $K$ (see (25))
by $S_0$.  Since we are looking for the correction with respect to
$\rho/d$ we substitute $\Delta_i$ by $\delta_i$ (see (15)). Thus the
quantity under consideration is \be
A'_{12}=<u^1|S^{-1}_0\delta_2S^{-1}_0\delta_1S^{-1}_0|u^2>.
\ee
By chirality arguments $\delta_i(i=1,2)$ should not contain
contributions proportional to $(1+\gamma_5)/2$ (formally such
contributions are found in $(S'_i-G_0)$, see (8)). Next we write
$$
A'_{12}\sim \int dz\{
\int dx u^{1+}(x) \partial_x\delta_2
(x-z)\}\partial_z\{
\int dy \delta_1(z-y)\partial_y u^2(y-d)\}
\equiv
$$
\be
\equiv
\int dz J_x(z)\partial_zJ_y(z).
\ee

Consider the  contribution from the region $x\leq\rho, z\simeq
d\pm\rho, y\simeq d\pm \rho$. In this region
$u^{1+}\partial_x\sim \rho^{-3}$. To estimate $\delta_2
(x-z)$ consider equation (8) for $S'_2$
with all arguments shifted by $d$ since $S'_2$ corresponds to $\bar
I$ placed at the distance $d$ from the origin. It is easy to conclude
that the leading contribution to $\delta_2(x-z)$ stems from the first
term of the second line of (8) and $\delta_2(x-z)\sim
G_0(x-z)\sim d^{-3}$. Therefore $J_x(z)\sim
\rho^4\rho^{-3}d^{-3}\sim \rho d^{-3}$. In $J_y(z) $ one
has $\partial_yu^2(y-d)\sim\rho^{-3}$. The estimate of
$\delta_1(z-y)$ deserves some attention. It also
receives the main contribution from the first term of
the second line of (18) but this time an additional
small parameter arises:
$$
\delta_1(z-y)\sim G_0(z-y)
[(\pi_x\pi_y)^{-1/2}(1+\frac{\rho^2}{x^2y^2}\tau_x\tau_y^+)-1]\sim
$$
\be
\sim G_0(z-y) \rho^3d^{-3}\sim (z-y)^{-3}\rho^3d^{-3}.
\ee
Therefore
$$
J_y(z)\sim \int dy(z-y)^{-3}d^{-3},
$$
where at the final step the integration over $y$ will
yield the $\rho^4$ factor. Now $\partial_z$ in (40)
acting on $(z-y)^{-3}$ in (41) will result in
$(z-y)^{-4}\sim \rho^{-4}$ since both $z$ and $y$ are
$\sim (d\pm\rho)$. Substituting everything back into
(40) we obtain
\be
A'_{12}\sim \rho^5 d^{-6}.
\ee
Since $V_{12}$ itself is $\sim \rho^2 d^{-3}$, the
correction to it is of the order $0((\rho/d)^3)$.

Now we turn to $m\rho$ corrections. The leading one
arises from $\alpha$ and $\beta$. Consider
\be
\alpha'=<u^2|S^{-1}_0\Delta_1S^{-1}_0|u^2>.
\ee

Chirality arguments are important here. If we substitute
both $S_0^{-1}$ by $G_0^{-1}$ and $\Delta_1$ by
$\delta_1$ we get a zero. Therefore one should consider
contributions arising from "unslashed" piece of any one
of the three operators inside (43) and "slashed" pieces
of the two others:
$$
\alpha'\simeq -im <u^2|\delta_1G^{-1}_0|u^2>
-im <u^2|G^{-1}_0\delta_1|u^2>+
$$
\be
+im <u^2|G^{-1}_0(S^{'2}_1-G^2_0)G^{-1}_0|u^2>\simeq
im <u^2|G^{-1}_0\delta^2_1G^{-1}_0|u^2>.
\ee
The last matrix element may be represented as
$$
\alpha'\sim m \int dz\{ \int dx u^{2+}
(x-d)\partial_x\delta_1(x-z)\}\times
$$
\be
\times \{\int dy \delta_1(z-y) \partial_y u^2(y-d)\}\equiv m\int dz
L_x(z) L_y(z).
\ee
First consider the contribution from the "far" region $x,y,z\sim d\pm
\rho$. Then $u^{2+}(x-d) \partial_x\sim \rho^{-3},\delta_1(x-z)\sim
d^{-3}$ (see (41)), $L_x(z) \sim \rho^4\rho^{-3} d^{-3} \sim  \rho
d^{-3}$.

The same estimate holds for $L_y(z)$ and thus
\be
\alpha'\sim m\rho^4(\rho d^{-3})^2\sim m\rho^6d^{-6}.
\ee
One may similarly verify  the estimate (46) for the "mixed" region
$x,y\sim d\pm\rho, z\leq \rho$. Therefore the $m^2$ term in the
effective action (38) acquires a correction of the order
$0(\rho^6/d^6)$.

Thus, we conclude that the leading--order corrections to the
effective action (38) are given by
\be
\tilde \Gamma_{eff} = N_f ln\{
m^2[1+0(\rho^6/d^6)]+V_{12}V_{21}+(V_{12}+V_{21}) 0(\rho^5/d^6)\},
\ee
 and we remind  that $V_{ij}\sim \rho^2/d^3$. We see that the
 corrections to $V_{12}V_{21}$ in the effective action are of order
 $\rho^3/d^3\sim (1/3)^3$. A careful look into various normalization
 factors, powers of $\pi$, etc. enhances this result by an order of
 magnitude. Here however a caution is needed since we are not  in  a
 position to account for possible numerical factors resulting from
 the integration procedure.

 Finally we wish to add one more remark on our results (38) and (47)
 for $\tilde \Gamma_{eff}$. The value of the overlap $V_{ij}$ being
 proportional to $\rho^2/d^3$ depends also on the relative color
 orientation of   $I$ and $\bar I$. Maximal attraction is reached
 under the condition that the relative color orientation vector
 $u_\mu=tr(\Omega_{\bar I}\tau_{\mu}\Omega^+_I)$ is parallel to
 $d_{\mu}$ [14]. On the other hand the matrix element $A_{ij}$ has a
 different dependence on the color orientation. Therefore the most
 favorable color orientations for $V_{ij}V_{ji}$ and $V_{ij}A_{ji}$
 entering into $\tilde \Gamma_{eff}$ are different.
 The problem of the polarization of the $I\bar I$ molecule with
 higher modes included will be discussed elsewhere.

 \section{ Summary}

In this parer  we have obtained exact solutions for the quark propagator and
the  effective  action in the instanton--anti--instanton mocecule. Up to now
a consistent treatment of the non--zero--modes has been lacking. Possible
applications of the results might be twofold. The first one concerns the
scenarious of the chiral symmetry breaking and restoration involving $I\bar
I$ molecules. Detailed properties of the fermion determinant are relevant for
this   subject -- see e.g. [11-15]. Another posible application involves
quark interactions induced by $I\bar I$ molecules and hadron
correlators [14]. The corresponding studies were so far based on the
quark propagator with non--zero-- modes  neglected. In such
approximation the propagator exhibits the symmetry properties which
are traceable in meson and baryon spectra. We have shown the
relevance of non--zero--modes in the quark propagator. To which
extent would higher modes affect the predicted properties of the
spectrum? The resulrs of the present paper allow to study this
problem. The work is in progress now.

\section{Acknowledgments}

We wish to thank V.A.Rubakov for enlightening discussion which motivated our
work. Fruitfull discussions with N.O.Agasyan and A.V.Smilga and
especially Yu.A.Simonov are greatly acknowledged. The work was
supported by the Russian Fund for Fundamental Research grants
97-02-16406, 96-02-19184, 96-15-96740 and RFFR-DFG 96-02-00088G.

\end{document}